\documentclass[aps,pra,twocolumn,showpacs]{revtex4-1}

\usepackage{graphicx}% Include figure files
\usepackage{dcolumn}% Align table columns on decimal point
\usepackage{bm}% bold math
\usepackage[version=3]{mhchem}
\usepackage{color}

\begin{document}

\title{Hybrid density functional study of optically active Er$^{3+}$ centers in GaN}
\author{Khang Hoang}
\email[E-mail: ]{khang.hoang@ndsu.edu}
%\homepage[]{Your web page}
%\thanks{}
%\altaffiliation{}
\affiliation{Center for Computationally Assisted Science and Technology, North Dakota State University, Fargo, ND 58108, USA.}

\date{\today}

\begin{abstract}

Understanding the luminescence of GaN doped with erbium (Er) requires a detailed knowledge of the interaction between the rare-earth dopant and the nitride host, including intrinsic defects and other impurities that may be present in the host material. We address this problem through a first-principles hybrid density functional study of the structure, energetics, and transition levels of the Er impurity and its complexes with N and Ga vacancies, substitutional C and O impurities, and H interstitials in wurtzite GaN. We find that, in the interior of the material, Er$_{\rm Ga}$ is the dominant Er$^{3+}$ center with a formation energy of 1.55 eV; Er$_{\rm Ga}$-$V_{\rm N}$ possesses a deep donor level at 0.61 eV which can assist in the transfer of energy to the 4\textit{f}-electron core. Multiple optically active Er$^{3+}$ centers are possible in Er-doped GaN.

\end{abstract}

% insert suggested PACS numbers in braces on next line
\pacs{}
% insert suggested keywords - APS authors don't need to do this
%\keywords{}

%\maketitle must follow title, authors, abstract, \pacs, and \keywords
\maketitle

% body of paper here - Use proper section commands
% References should be done using the \cite, \ref, and \label commands

{\bf 1 Introduction} Rare-earth dopants in semiconductors have received great attention as they lead to sharp intra-\textit{f} shell optical transitions which are desirable for many optoelectronic applications. When optically excited, Er-doped GaN, for example, will emit visible light in a few narrow bands in the green \cite{ODonnell2006}. The Er center can be excited by a direct absorption of energy into the 4\textit{f}-electron core (resonant excitation) or indirectly by energy transfer from the host (non-resonant excitation). The latter mechanism is believed to be mediated by Er-related defects in which the presence of defect levels in the host band gap, acting as carrier traps, may have a crucial role. It has been reported that multiple optically active Er$^{3+}$ centers are present in the material; one center type may only be excited by resonant excitation while others are excited via non-resonant excitation \cite{Kim1997,Przybylinska2001,Braud2003,Dierolf2004}. Song et al.~observe different energy levels, at 0.188, 0.300, 0.410, and 0.600 eV below the conduction-band minimum (CBM), presumably corresponding to different intrinsic and Er-related defects, in deep-level transient spectroscopy (DLTS) experiments on n-type GaN implanted with Er \cite{Song2007}. However, despite much effort, there is still no definitive assignment of these levels and the observed Er$^{3+}$ centers to specific defect configurations.

Computational studies of Er impurities in GaN have been carried out by several groups, e.g., using the local-density approximation (LDA) or self-interaction corrected LDA within density-functional theory (DFT) or the so-called LDA$+U$ within a DFT-based tight-binding approach \cite{Filhol2004,Svane2006,Sanna2009}. Filhol et al., for example, report that substitutional Er at the Ga site (Er$_{\rm Ga}$) has a formation energy of 0.84 eV \cite{Filhol2004}, while Sanna et al.~find a much higher value ($>$6 eV) \cite{Sanna2009}. These values are either much lower or higher than the experimental formation energy of 1.8$\pm$0.2 eV reported by Ugolini et al.~for optically active Er$^{3+}$ centers in GaN~\cite{Ugolini2012}. We herein present a first-principles study of the Er impurity and its complexes with nitrogen and gallium vacancies, substitutional carbon and oxygen impurities, and hydrogen interstitials, using a hybrid functional approach. Although there are still limitations in the particular case of strongly correlated electrons, the method has been shown to be promising in computational studies of defects in solids \cite{Freysoldt2014RMP}. Carbon, oxygen, and hydrogen are common unintentional dopants in GaN. In light of our results for defect structures, formation energies, and transition levels, possible optically active Er$^{3+}$ centers are identified.

{\bf 2 Methods} Our calculations are based on DFT with the Heyd-Scuseria-Ernzerhof (HSE) functional \cite{heyd:8207}, as implemented in VASP \cite{VASP2}. The Hartree-Fock mixing parameter is set to 0.31, resulting in a band gap of 3.53 eV for GaN, very close to the experimental value ($\sim$3.5 eV). The defects are simulated using a 96-atom supercell, a plane-wave basis-set cutoff of 400 eV, and a 2$\times$2$\times$2 $k$-point mesh for the integrations over the Brillouin zone. In defect calculations, the lattice parameters are fixed to the calculated bulk values but all the internal coordinates are relaxed; spin polarization is included. The Ga 3\textit{d} and Er 4\textit{f} electrons are treated as core states within the projector augmented wave method \cite{PAW1,PAW2}. Lyons et al.~reports that the inclusion of the Ga semicore \textit{d} electrons in the valence has negligible effects on the formation energies and transition levels of defects in GaN~\cite{Lyons2014PRB}. Our HSE calculations of ErGa$_{47}$N$_{48}$ in which the Er 4\textit{f} electrons are explicitly considered show that there are no \textit{f} states in the band gap or near the valence-band maximum (VBM) and CBM, indicating that the explicit inclusion of the \textit{f} electrons in the valence would have negligible effects on the energetics of Er-related defects in GaN. Note that un-spin-polarized DFT+$U$ results reported by Wang et al.~\cite{Wang2012} also show the VBM and CBM are predominantly bulk GaN states.

The likelihood of a intrinsic defect, impurity, or defect complex X in charge state $q$ being incorporated into a crystal is characterized by its formation energy, defined as
\begin{align}\label{eq:eform}
E^f({\mathrm{X}}^q)&=&E_{\mathrm{tot}}({\mathrm{X}}^q)-E_{\mathrm{tot}}({\mathrm{bulk}}) -\sum_{i}{n_i\mu_i} \\ %
\nonumber &&+~q(E_{\mathrm{v}}+\mu_{e})+ \Delta^q ,
\end{align}
where $E_{\mathrm{tot}}(\mathrm{X}^{q})$ and $E_{\mathrm{tot}}(\mathrm{bulk})$ are the total energies of the defect and bulk supercells; $n_{i}$ is the number of atoms of species $i$ that have been added ($n_{i}>0$) or removed ($n_{i}<0$) to form the defect; $\mu_{i}$ is the atomic chemical potential, representing the energy of the reservoir with which atoms are being exchanged. The chemical potentials of Ga, C, N, and H are referenced to the total energy per atom of bulk Ga, bulk C (diamond), N$_2$ and H$_2$ at 0 K, while those of Er and O are assumed to be limited by the formation of ErN and Ga$_2$O$_{3}$. The formation enthalpies ($\Delta H$, at 0 K) are calculated to be $-$1.26, $-$4.00, and $-$10.07 eV for GaN, ErN, and Ga$_2$O$_{3}$, in agreement with the experimental standard formation enthalpies of $-$1.63$\pm$0.17 \cite{Ranade2000}, $-$3.71$\pm$0.23 \cite{Kordis1977}, and $-$11.20 eV \cite{Zinkevich2004}. The chemical potentials of Ga and N can vary over a range determined by the formation enthalpy of GaN such that $\mu_{\rm Ga}+\mu_{\rm N} = \Delta H({\rm GaN})$; $\mu_{\rm Ga} = 0$ ($\mu_{\rm N} = 0 $) corresponds to Ga-rich (N-rich) condition. $\mu_{e}$ is the electronic chemical potential, i.e., the Fermi level, representing the energy of the electron reservoir, referenced to the VBM in the bulk ($E_{\mathrm{v}}$). $\Delta^q$ is the correction term to align the electrostatic potentials of the bulk and defect supercells and to account for finite-size effects on the total energies of charged defects~\cite{Freysoldt}. %Freysoldt11

From defect formation energies, one can calculate thermodynamic transition levels (not to be confused with optical transition levels) between charge states $q$ and $q'$ of a defect, $\epsilon(q/q')$, defined as the Fermi-level position at which the formation energy of the defect in charge state $q$ is equal to that in charge state $q'$ \cite{Freysoldt2014RMP}. $\epsilon(q/q')$ is independent of the atomic chemical potentials. This level would be observed in, e.g., DLTS, where the defect in the final charge state fully relaxes to its equilibrium configuration after the transition.  

\begin{figure}[t]%
\vspace{0.2cm}
\includegraphics*[width=\linewidth]{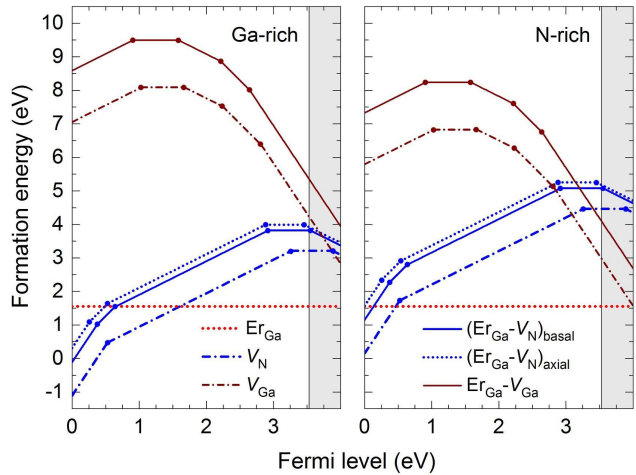}
\caption{Formation energies of $V_{\rm N}$, $V_{\rm Ga}$, Er$_{\rm Ga}$, and Er$_{\rm Ga}$-$V_{\rm N}$ and Er$_{\rm Ga}$-$V_{\rm Ga}$ complexes in GaN, plotted as a function of Fermi level, under Ga-rich and N-rich conditions. Fermi levels above the CBM (at 3.53 eV) are indicated by the shaded area.}
\label{fig;fe1}
\end{figure}

{\bf 3 Results and discussion} In bulk GaN, each Ga atom is coordinated with four N atoms: one along the $c$-axis with the Ga$-$N bond length calculated to be of 1.958 {\AA} and three in the basal plane with the Ga$-$N bond length of 1.952 {\AA}. Figure~\ref{fig;fe1} shows the calculated formation energies of the nitrogen vacancy ($V_{\rm N}$), gallium vacancy ($V_{\rm Ga}$), Er$_{\rm Ga}$, and complexes between Er$_{\rm Ga}$ and the vacancies. $V_{\rm N}$ is found to be stable in the 3+, +, and neutral charge states; the axial and basal positions of $V_{\rm N}$ are degenerate in energy. This vacancy introduces two transition levels in the band gap: one at 0.53 eV above the VBM corresponding to the transition between the 3+ and + charge states and the other at 0.27 eV below the CBM corresponding to the transition between the + and neutral states, in good agreement with those reported by Yan et al.~\cite{Yan2012APL}. $V_{\rm Ga}$ is stable in all charge states from 3$-$ to +. The (+/0) transition level occurs at 1.03 eV, followed by the (0/$-$) level at 1.67 eV, the ($-$/2$-$) level at 2.24 eV, and the (2$-$/3$-$) level at 2.82 eV above the VBM, in agreement with Lyons et al.~\cite{Lyons2015}.   

We find that Er$_{\rm Ga}$ is stable only in the neutral state, {\it cf.}~Fig.~\ref{fig;fe1}. This is also observed in our calculations where Er 4\textit{f} electrons are included in the valence. The Er$^{3+}$ ion does not only introduce an outward relaxation of the neighboring N ions, but also distort the local lattice environment, making the trigonal distortion much more pronounced with the Er$-$N bond length along the $c$-axis is $\sim$0.03 {\AA} longer than the other bonds; the Er$-$N bond lengths are calculated to be 2.141, 2.143, and 2.148 {\AA} in the basal plane and 2.169 {\AA} along the $c$-axis. Er$^{3+}$ also goes off-center by moving toward the basal N plane by 0.08 {\AA}. These results are in contrast to previous work which argues that Er$^{3+}$ remains ``exactly on-center'' and even reduces the difference in the axial and basal bond lengths \cite{Sanna2009,Konopka2011}. The formation energy of Er$_{\rm Ga}^0$ is calculated to be 1.55 eV and remains unchanged in going from Ga-rich to N-rich conditions, as it depends on the atomic chemical potentials only through the difference in the formation enthalpies of GaN and ErN. This value corresponds to, e.g., a concentration of 5$\times$10$^{16}$ cm$^{-3}$ under thermal equilibrium at 1040$^\circ$C. In the absence of defects and other impurities in its close proximity, Er$_{\rm Ga}$ would thus be electrically inert. Still, intra-4\textit{f} shell optical transitions can occur in these isolated Er$^{3+}$ centers as the local lattice distortion mentioned above breaks the inversion symmetry, thus relaxing the parity selection rules.

\begin{figure}[t]%
\vspace{0.2cm}
\includegraphics*[width=\linewidth]{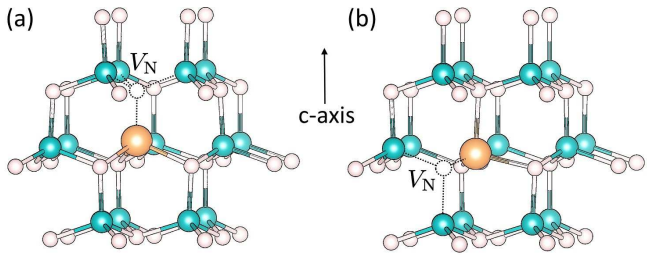}
\caption{Structure of (Er$_{\rm Ga}$-$V_{\rm N}$)$^0$: (a) axial and (b) basal configurations. Large circles are Er, medium Ga, and small N.}
\label{fig;struct}
\end{figure}

Regarding Er$_{\rm Ga}$-$V_{\rm N}$, there are two configurations corresponding to different positions of $V_{\rm N}$ in the complex: ``axial'' or ``basal'' as shown in Fig.~\ref{fig;struct}. They both have large local lattice distortions; Er$^{3+}$ moves off-center toward the vacancy by 0.22 {\AA} (0.20 {\AA}) and the distance between the two components in (Er$_{\rm Ga}$-$V_{\rm N}$)$^0$ is 1.73 {\AA} (1.77 {\AA}) in the basal (axial) configurations. The basal configuration has a lower formation energy, as seen in Fig.~\ref{fig;fe1}. In the following, we refer to a defect complex only in its lowest-energy configuration. Note that the local distortion is also different for different charge states. In (Er$_{\rm Ga}$-$V_{\rm N}$)$^{3+}$, for example, the distance between the two constituents is 1.66 {\AA}; Er$^{3+}$ thus moves toward the vacancy by 0.29 {\AA}. The binding energy of (Er$_{\rm Ga}$-$V_{\rm N}$)$^q$, with respect to its isolated constituents Er$_{\rm Ga}^0$ and $V_{\rm N}^q$, is 0.94, 0.60, 0.74, 0.56, or 1.28 eV for $q$ = 0, +, 2+, 3+, or $-$, respectively. The binding energy is thus relatively high (0.94 eV) in n-type GaN where Er$_{\rm Ga}$-$V_{\rm N}$ is most stable in the neutral state, {\it cf.}~Fig.~\ref{fig;fe1}, suggesting that a significant portion of $V_{\rm N}$ can exist in form of Er$_{\rm Ga}$-$V_{\rm N}$  \cite{Filhol2004}. This defect complex has several deep levels in the band gap, corresponding to the (3+/2+) level at 0.38 eV and the (2+/+) level at 0.64 eV above the VBM, and the (+/0) level at 0.61 eV below the CBM. The (0/$-$) level is only 0.02 eV above the CBM, {\it cf.}~Fig.~\ref{fig;fe1}. The deep donor level at 0.61 eV below the CBM can act as a carrier trap; an electron captured here can subsequently recombine nonradiatively with a free hole from the valence band or a hole at some acceptor level and transfer energy to the 4\textit{f} shell of the Er$^{3+}$ ion. We note that the concentration of Er$_{\rm Ga}$-$V_{\rm N}$ should be much smaller than that of the isolated Er$_{\rm Ga}$; e.g., assuming n-type GaN and extreme Ga-rich condition, the formation energy of the complex is 3.82 eV, corresponding to an equilibrium concentration of only 3$\times$10$^8$ cm$^{-3}$ at 1040$^\circ$C. However, the cross section for electron capture associated with the complex is expected to be larger, as indicated in experiments~\cite{Braud2003}. Note that the concentration of certain defects could be much higher near the surface or if the material is prepared via a non-equilibrium process.  

For (Er$_{\rm Ga}$-$V_{\rm Ga}$)$^q$, we find a binding energy of 0.44, 0.26, 0.24, 0.15, and 0.02 eV with respect to Er$_{\rm Ga}^0$ and $V_{\rm Ga}^q$ for $q$ = 3$-$, 2$-$, $-$, 0, or +, respectively. These values are thus much smaller than those in the case of (Er$_{\rm Ga}$-$V_{\rm N}$)$^q$, mainly due to the larger distance between Er$_{\rm Ga}$ and $V_{\rm Ga}$. For example, the distance between the two components in (Er$_{\rm Ga}$-$V_{\rm Ga}$)$^{3-}$, the most stable charge state in n-type GaN, is 2.75 {\AA}. Er$_{\rm Ga}$-$V_{\rm Ga}$ induces several deep levels in the band gap: the (+/0), (0/$-$), ($-$/2$-$), and (2$-$/3$-$) levels occur at 0.91, 1.58, 2.22, and 2.65 eV above the VBM, respectively. Some of these levels may act as carrier traps and thus assist in the transfer of energy to the Er 4\textit{f} shell. However, given the complex's relatively low binding energy and high formation energy (e.g., compared to Er$_{\rm Ga}$-$V_{\rm N}$, except at the Fermi levels very close to the CBM and under extreme N-rich condition, {\it cf.}~Fig.~\ref{fig;fe1}), the role of Er$_{\rm Ga}$-$V_{\rm Ga}$ in the Er-related luminescence is still not clear at this point.       
	
\begin{figure}[t]%
\vspace{0.2cm}
\includegraphics*[width=\linewidth]{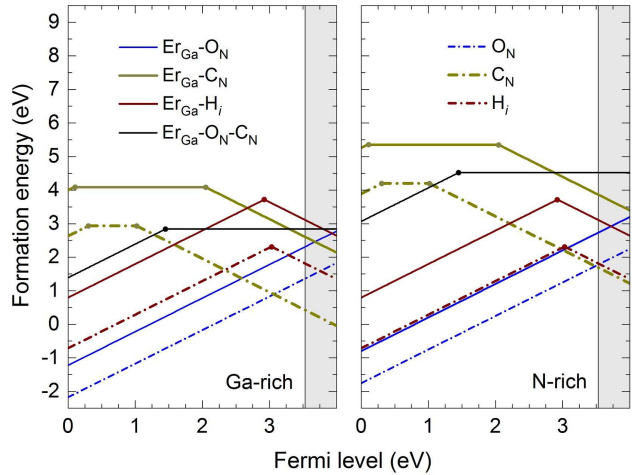}
\caption{Formation energies of O$_{\rm N}$, C$_{\rm N}$, and H$_i$, and their complexes with Er$_{\rm Ga}$ in GaN under Ga-rich and N-rich conditions. Fermi levels above the CBM are indicated by the shaded area.}
\label{fig;fe2}
\end{figure}

Figure~\ref{fig;fe2} shows the calculated formation energies of the substitutional oxygen (O$_{\rm N}$) and carbon (C$_{\rm N}$) at the N site and hydrogen interstitial (H$_i$), and their complexes with Er$_{\rm Ga}$. O$_{\rm N}$ is found to be a shallow donor. C$_{\rm N}$ induces two levels in the band gap: the (+/0) level at 0.31 eV and the ($-$/0) level at 1.02 eV above the VBM. H$_i$ is amphoteric, being positively charged in p-type and negatively charged in n-type GaN; the (+/$-$) level occurs at 3.03 eV above the VBM. Our results for O$_{\rm N}$, C$_{\rm N}$, and H$_i$ are thus in good agreement with those reported previously~\cite{Lyons2014PRB,Gordon2014,Lyons2012}.

The energy landscape of the complexes between Er$_{\rm Ga}$ and O$_{\rm N}$, C$_{\rm N}$, and H$_i$ resembles that of the isolated non-Er impurities, {\it cf.}~Fig.~\ref{fig;fe2}. Like O$_{\rm N}$, Er$_{\rm Ga}$-O$_{\rm N}$ is also a shallow donor; the complex in its + charge state has a binding energy of 0.60 eV with respect to Er$_{\rm Ga}^0$ and O$_{\rm N}^+$. Since there are no in-gap levels associated with Er$_{\rm Ga}$-O$_{\rm N}$, this complex is unlikely to be effective in assisting the energy transfer to the Er 4\textit{f} shell. Er$_{\rm Ga}$-C$_{\rm N}$ induces two levels in the band gap: the (+/0) level at 0.11 eV and the (0/$-$) level at 2.05 eV. However, the most stable charge state of the complex in n-type GaN, (Er$_{\rm Ga}$-C$_{\rm N}$)$^-$, has a negative binding energy of $-$0.63 eV and is thus unstable with respect to Er$_{\rm Ga}^0$ and C$_{\rm N}^-$. The binding energy of other charge states is also quite low: 0.40 or 0.20 eV for the neutral or $+$ state. We also consider the Er$_{\rm Ga}$-O$_{\rm N}$-C$_{\rm N}$ complex where O is at one of the basal N positions and C is at the axial position; the binding energy is 0.49 eV in its neutral state or 0.94 eV in the + state. The (+/0) level induced by this complex is, however, quite low at 1.46 eV above the VBM, {\it cf.}~Fig.~\ref{fig;fe2}, thus excluding its possible role in high-energy luminescent transitions. Finally, Er$_{\rm Ga}$-H$_i$ is found to have a binding energy of only 0.04 or 0.25 eV in the + or $-$ charge state, suggesting that it is unlikely to be stable as a defect complex.

In n-type GaN, $V_{\rm N}^0$ is spin-polarized with spin S = 1/2 and (Er$_{\rm Ga}$-$V_{\rm N}$)$^0$ has a total spin of 2 as seen in our calculations where the Er 4\textit{f} electrons are explicitly included; the $V_{\rm N}^-$ component of (Er$_{\rm Ga}$-$V_{\rm N}$)$^-$ has S = 1. Other defects relevant to n-type GaN are not spin-polarized, except the Er$_{\rm Ga}^0$ component which has spin S = 3/2. Experimentally, Ugolini et al.~obtain a formation energy of 1.8$\pm$0.2 eV for Er$^{3+}$ centers in Er-doped GaN prepared by MOCVD. These authors presume the centers to be Er$_{\rm Ga}$-$V_{\rm N}$~\cite{Ugolini2012}. We find that, given the much higher formation energies of both $V_{\rm N}$ and the complex in n-type GaN, {\it cf.}~Fig.~\ref{fig;fe1}, those Er$^{3+}$ centers should instead be identified with isolated Er$_{\rm Ga}$ which has a calculated formation energy of 1.55 eV. Regarding $V_{\rm N}$, the (+/0) transition level can be identified with the DLTS level at 0.270 eV in undoped GaN or 0.300 eV in Er-implanted GaN observed in experiments~\cite{Song2007}. Among the remaining DLTS levels reported by Song et al., the level at 0.600 eV can be identified with the (+/0) level of Er$_{\rm Ga}$-$V_{\rm N}$; other levels at 0.188 eV (often mentioned in the literature as being related to Er$_{\rm Ga}$-$V_{\rm N}$) and 0.410 eV are likely to involve some other defect structures including intrinsic and/or Er-related defects near the surface. Further studies are needed to clarify the defect physics of Er-doped GaN. 

{\bf 4 Summary and conclusions} In summary, we find that isolated Er$_{\rm Ga}$ is the dominant Er$^{3+}$ center, at least in n-type GaN where it has the lowest formation energy among all Er-related defects. This impurity possesses a local lattice distortion that can help relax the selection rules, thus allowing for intra-4\textit{f} transitions to occur in the Er$^{3+}$ ion. Er$_{\rm Ga}$-$V_{\rm N}$ is identified as an efficient defect-related Er center for band-to-band (non-resonant) excitation. This complex has the highest binding energy, shortest distance between the defect and Er, and a deep trap at 0.61 eV below the CBM. Er$_{\rm Ga}$-$V_{\rm Ga}$, in principle, can also act as an assistant for the transfer of energy to the Er 4\textit{f} shell. Multiple luminescent centers with, e.g., different Stark splittings, luminescence decay dynamics, and excitation cross sections are thus possible in real Er-doped GaN samples.         

{\bf Acknowledgments} We thank J.~L.~Lyons, J.~B.~Varley, Q.~Jan, and N.~Q.~Vinh for very useful discussions. This work was supported through the U.S.~Department of Energy Grant No.~DE-SC0001717 and by the Center for Computationally Assisted Science and Technology at North Dakota State University. 

\providecommand{\WileyBibTextsc}{}
\let\textsc\WileyBibTextsc
\providecommand{\othercit}{}
\providecommand{\jr}[1]{#1}
\providecommand{\etal}{~et~al.}

\end{document}